\documentclass[final,3p,times]{elsarticle}
\usepackage{amssymb}
\usepackage{amsmath}
\usepackage{slashed}
\journal{Journal of Subatomic Particles and Cosmology}

\begin{document}

\begin{frontmatter}



\title{Remembering Brown-Rho scaling}


\author[label1]{Su Houng Lee} %% Author name
\ead{suhoung@yonsei.ac.kr}

\affiliation[label1]{organization={Department of Physics and Institute of Physics and Applied Physics, Yonsei University},%Department and Organization
%           addressline={50 Yonseiro},
           city={Seoul},
           postcode={03722},
           country={Korea}}

%% Abstract
\begin{abstract}
The properties of hadrons in dense matter have been a core interest of mine since I entered graduate school at Stony Brook in 1984. This interest was shaped both by the strong research activity on the subject at Stony Brook at the time and by the influence of Mannque Rho, who was a frequent visitor there. In this memoir dedicated to Mannque, I describe a selection of his works that sparked my interest in the subject and present my current perspective on what Mannque, together with Gerry Brown, sought to emphasize in the often-misinterpreted Brown--Rho scaling. In light of recent developments, I also discuss refinements of Brown--Rho scaling for mesons and baryons, associated with spontaneous chiral symmetry breaking and $U_A(1)$-breaking effects.
\end{abstract}

%% Keywords

\end{frontmatter}



\section{Introduction}
\label{sec1}
%% Labels are used to cross-reference an item using \ref command.

I first met Mannque Rho in 1985 when I joined the nuclear theory group at Stony Brook as a graduate student. Although my thesis advisor was Hans Hansson, since I belonged to the nuclear theory group headed by Gerry Brown and Mannque was a frequent visitor, I had many opportunities to interact with both Gerry and Mannque. I learned a great deal from them and came to regard them not only as teachers but also as mentors who profoundly influenced my development as a physicist. Furthermore, because Mannque was originally from Korea, frequently visited Korea, and we shared many common interests, we remained in close contact over the years. As recently as the past few years, I invited him to contribute review articles to the Journal of Subatomic Particles and Cosmology \cite{Rho:2024ihu,Rho:2025rqf}, of which I am the  Editor-in-Chief. I was particularly pleased when Mannque's most recent article was selected as the Editor's Choice of our journal \cite{Rho:2024ihu}.

Mannque Rho has many seminal works in nuclear theory and effective field theory, some of which are highlighted in the other articles in this contribution.  A common denominator to all his work is chiral symmetry.  In this article, I would like to highlight few papers that lead to my own interest and his work on Brown-Rho scaling.  

\section{Cleaning of the vacuum}

Chiral symmetry breaking is one of the important non-perturbative effect in QCD.  At the same time, the spontaneously broken symmetry is expected to be restored at high temperature and/or nuclear density.  The high temperature needed for the restoration is certainly reached at the initial stages of the ultra relativistic heavy ion collision. Unfortunately, after the collision, the system will cool down and go through the hadronic phase after which most of the signatures will be washed out.  Identifying the signatures of chiral symmetry restoration in heavy ion collision remains a challenge.  Possible signatures are discussed in Ref.\cite{Lee:2025nix}.  On the other hand, it is known that chiral symmetry will be partly restored even in the nuclear saturation density.  Hence, a heavy nucleus provides a stable environment in which one can observe precursor phenomena of chiral symmetry restoration.   

The leading density dependence of the light quark condensate in nuclear matter
can be obtained in the linear density approximation,
\begin{equation}
\frac{\langle\bar q q\rangle_\rho}
{\langle\bar q q\rangle_0}
=
1
+ c \rho ,  ~~~ {\rm where }~~~c= \frac{\langle p| \bar{q}q | p \rangle }{\langle\bar q q \rangle_0}= -
\frac{\sigma_{\pi N}}
{f_\pi^2m_\pi^2}. 
\label{condensate-linear}
\end{equation}
Here $\langle p| \bar{q}q |p \rangle$ is the proton expectation value of the chiral order parameter
Using the lates value of $\sigma_{\pi N} \sim $ 43.7 MeV, $f_\pi \sim$ 92.3 MeV. $m_\pi \sim $138 MeV, the change will be more than 30\% reduction at nuclear matter density $\rho_0 \simeq 0.16 $fm$^{-3}$.  

\begin{figure}[t]
\centering%% For centre alignment of image.
\includegraphics[scale=0.5]{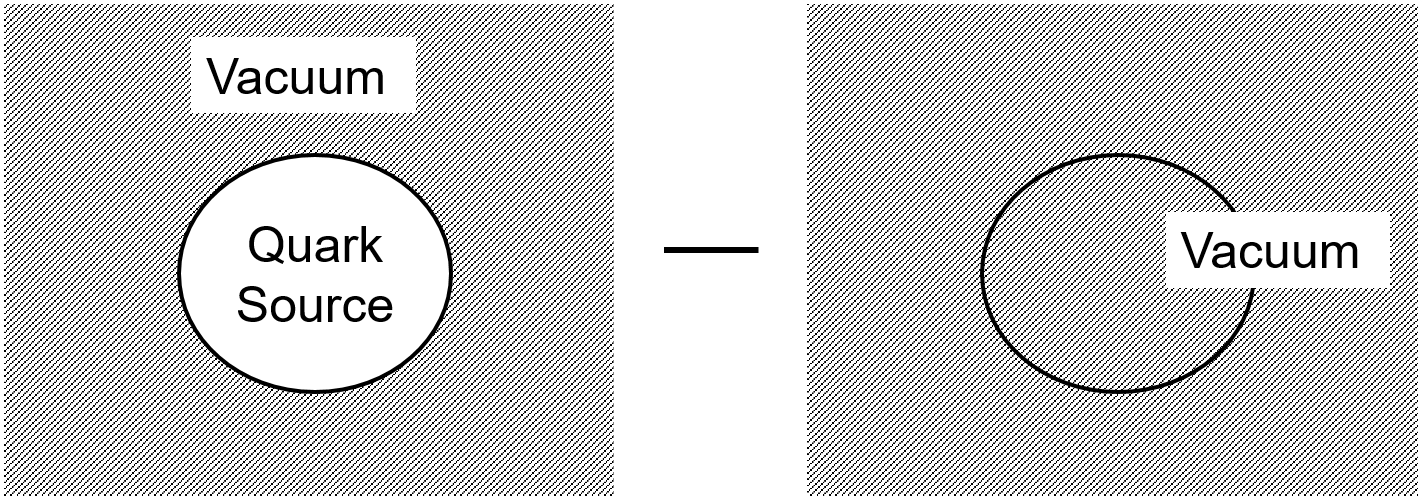}
%% Use \caption command for figure caption and label.
\caption{A pictorial representation of Eq.~\eqref{vacuum-cleaning}.  The proton is depicted as the bubble.}\label{fig1}
%% https://en.wikibooks.org/wiki/LaTeX/Importing_Graphics#Importing_external_graphics
\end{figure}

Although the formula given Eq.~\eqref{condensate-linear}  is given in Ref.\cite{Drukarev:1988kd}, there is an interesting earlier work by Mannque Rho and his collaborators that shed light into the origin of the reduction due to the change in the vacuum inside the proton\cite{Brown:1987hj}.  The paper addresses kaon condensation by considering the proton expectation value of the mass term in the QCD Lagrangian. To simplify the discussion, let us consider the proton expectation value of the light-flavor contributions to the mass terms in the formula given in Ref.~\cite{Brown:1987hj}.
\begin{align}
 \langle p |H_{SB}| p \rangle   
 = \int dV  \langle p |H_{SB}| p \rangle_{\bf bubble}   
 -\int dV  \langle 0 |H_{SB}| 0 \rangle_{\bf vacuum}  
\label{vacuum-cleaning}
\end{align}
where $H_{SB}=m_u \bar{u}u+m_d \bar{d} d$, and the proton state is normalized as $ \langle p | p' \rangle= (2 \pi)^3 \delta^3 (p-p')$.
A diagrammatic representation of Eq.~\eqref{vacuum-cleaning} is shown in Fig.~\ref{fig1}.

Now in Ref.~\cite{Brown:1987hj}, the authors introduced the concept of vacuum cleaning, where the proton is assumed to be a spherical MIT bag in which chiral symmetry is restored inside the bag.  Hence,  one takes $\langle p |H_{SB}| p \rangle_{\bf bubble}  =0$ and $V=\frac{4 \pi}{3}R^3$.  Then the right hand side of Eq.\eqref{vacuum-cleaning}
can be approximated by $ V \times \langle 0 |H_{SB}| 0 \rangle_{\bf vacuum}  $.  Using the estimates from chiral perturbation theory and QCD sum rules, the authors were able to show that the symmetry breaking term could indeed be explain with a reasonable bag size.  
To be specific, applying the same approach to the chiral condensate one finds that  $c$ given in Eq.~\eqref{condensate-linear} can be identified as the volume $V$, from which we find $R=0.79$ fm, which is  the typical size of an MIT bag\cite{Chodos:1974je}.  The relation can be summarized as follows;
\begin{align}
\langle p|\bar{q}q|p\rangle_{\bf bubble}=0
\quad\Longrightarrow\quad
c &\equiv
\frac{\langle p|\bar{q}q|p\rangle}
{\langle\bar{q}q\rangle_0}
=- V
\simeq - 2.07~\mathrm{fm}^3
\quad\Longrightarrow\quad
R\simeq 0.79~\mathrm{fm}.
\end{align}

If one interprets each proton as a chiral-restored bubble of volume $V\sim 2.07,\mathrm{fm}^3$, then at nuclear saturation density the fraction of volume occupied by such bubbles is $V\rho_0\simeq0.33$. This provides an intuitive geometrical interpretation of the roughly 33\% reduction of the chiral condensate obtained in the linear-density approximation.

Hence, Mannque knew that chiral symmetry restoration is substantially reduced in nuclear matter and was looking for ways to observe this effect.

%% Use \subsection commands to start a subsection.
\section{Quenching of $g_A$}
\label{subsec1}

One of Mannque's enduring research interests was the quenching of ($g_A$) in nuclear matter. In nuclear ($\beta$) decay, one can select pure or predominantly Gamow-Teller transitions, thereby probing the axial component of the weak current. Experimentally, the Gamow-Teller matrix elements extracted from nuclear decays are found to be systematically smaller than those calculated in the independent-particle shell model using the free-nucleon value of ($g_A$). This discrepancy, commonly referred to as the Gamow-Teller quenching problem, is often parametrized in terms of an effective axial coupling ($g_A^{\rm eff}<g_A^{\rm free}$). 

In one of his early papers~\cite{Oset:1979ue}, he argued that this quenching could arise from spin--isospin correlations in the nuclear medium, with the short-range interaction characterized by the Landau--Migdal parameter $g'$. These correlations modify the Gamow--Teller response, including through the coupling of nucleon particle--hole excitations to $\Delta$-hole configurations, thereby reducing the strength of the low-energy Gamow--Teller transition. Using a phenomenological value of $g'$, he showed that this suppression could account for the observed reduction in the Gamow--Teller strength, conventionally expressed in terms of an effective axial coupling $g_A^{\rm eff}\sim 1$. This work exemplified Mannque's broader interest in understanding complex nuclear phenomena in terms of a small number of parameters characterizing effective theories.

It is important to emphasize that the quenching of $g_A$ does not necessarily imply a modification of the fundamental axial coupling of a free nucleon. Rather, it provides a convenient way of parametrizing the effects of nuclear many-body correlations on the Gamow--Teller response. Mannque was nevertheless interested in whether such effective-medium modifications could be understood more directly as changes in the parameters of an effective Lagrangian.

This line of thought became more apparent in his later work~\cite{Rho:1985ay}, which can be viewed as a precursor to Brown--Rho scaling. At the time, considerable attention was devoted to the Skyrme model, in which the nucleon emerges as a topological soliton of an effective chiral Lagrangian. The hedgehog ansatz determines the soliton profile in terms of the parameters of the chiral Lagrangian, most notably $f_\pi$ and the Skyrme parameter $e$ associated with the stabilizing term. Since $g_A$ is determined by the axial current of the soliton, one can relate the  the nucleon size with the parameters
of the model
$r_{\rm Skyrmion} \propto \sqrt{g_A}/f_\pi $.  Mannque further argued, using chiral Ward identities, that the medium dependence of $g_A$ is closely connected to that of $f_\pi$~\cite{Rho:1985ay}.  Hence,  the quenching of $g_A$ together with the reduction of $f_\pi$ lead to the swelling of the nucleon in nuclear matter. Mannque further argued, that the physics of swelling of the nucleon in nuclear matter is the origin of the quenching of $g_A$.

Later, however, Mannque revisited the origin of the apparent quenching of $g_A$ and emphasized that the phenomenological value $g_A^{\rm eff}\sim1$ should primarily be understood as a consequence of many-body correlations at the Landau Fermi-liquid fixed point, rather than as a fundamental modification of the axial coupling of the nucleon~\cite{Rho:2025rqf}.
However, as we will see in the next section, his earlier work~\cite{Rho:1985ay} led him to realize that the in-medium modification of hadron masses provides a more direct manifestation of changes in the parameters of an effective Lagrangian associated with chiral symmetry breaking, an idea that became central to his development of Brown--Rho scaling~\cite{Brown:1991kk}.

\section{Bron-Rho scaling}

The starting point of the Brown--Rho paper in Ref.~\cite{Brown:1991kk} is the relation obtained in Ref.~\cite{Rho:1985ay} for the radius and mass of the nucleon,
\begin{align}
\langle r^2 \rangle &\sim \frac{g_A}{f_\pi^2},
~~~
M_N \sim g_A^{1/2} f_\pi.
\label{r2m}
\end{align}
Here, $g_A$ is introduced in place of the original Skyrme parameter $e$, through the profile of the soliton solution.

In the 1985 work, the medium dependence of $g_A$ and $f_\pi$ was interpreted in terms of a medium-modified Skyrmion, with the resulting increase in the nucleon size associated with the quenching of $g_A$. However, in the 1991 paper, he emphasized that the phenomenological quenching of $g_A$ is related primarily to nuclear many-body effects rather than directly to the chiral order parameter. He therefore took the vacuum value of $g_A$ in Eq.~\eqref{r2m} and shifted the emphasis toward the medium dependence of the parameters of an effective chiral Lagrangian.

Mannque's approach was rooted in effective field theory. The starting point was to introduce scale invariance into the Skyrme Lagrangian,
\begin{align}
{\cal L} &=
\frac{f_\pi^2}{4}
\left(\frac{\chi}{\chi_0}\right)^2
{\rm Tr}(\partial_\mu U \partial^\mu U^\dagger)
+\frac{1}{32e^2}
{\rm Tr}
\left[
U^\dagger \partial_\mu U,
U^\dagger \partial_\nu U
\right]^2
+{\cal L}(\chi).
\end{align}
Here, $U=e^{i\pi/f_\pi}$ and $\chi$ is the dilaton field associated with the scale anomaly in QCD, with
\begin{equation}
\chi_0 \propto \langle 0|G^2|0\rangle^{1/4},
\end{equation}
where $\chi_0$ is the vacuum expectation value of the dilaton field and is proportional to the fourth root of the gluon condensate, with the appropriate normalization.

The next assumption was the identification of the in-medium pion decay constant as
\begin{align}
f_\pi^* = f_\pi \frac{\chi^*}{\chi_0},
\label{fpi-chi}
\end{align}
and the identification
\begin{align}
\frac{\langle 0^* | \bar{q}q|0^* \rangle}
{\langle 0 | \bar{q}q|0 \rangle}
=
\left(\frac{\chi^*}{\chi_0}\right)^3,
\label{chiraltogluon}
\end{align}
where $\langle 0^* | \bar{q}q|0^* \rangle$ denotes the in-medium chiral condensate. This led to the corresponding scaling of the nucleon mass in the medium.

To formulate the scaling of the vector-meson mass, Brown and Rho adopted the hidden local symmetry formulation of the chiral Lagrangian, in which the vector mesons emerge as the gauge bosons of the hidden local symmetry~\cite{Bando:1987br}. This approach gives the vector-meson mass
\begin{align}
m_\rho^2 = a g^2 f_\pi^2,
\end{align}
where $g$ is the hidden-gauge coupling. For $a=2$, this relation reduces to the KSRF relation. Assuming that the hidden-gauge coupling and the parameter $a$ vary only weakly at low density, the vector-meson mass was predicted to scale approximately with $f_\pi$.

Combining similar arguments for the sigma meson, he arrived at the Brown--Rho scaling relation:
\begin{equation}
\frac{m_\sigma^*}{m_\sigma}
\simeq
\frac{m_N^*}{m_N}
\simeq
\frac{m_\rho^*}{m_\rho}
\simeq
\frac{m_\omega^*}{m_\omega}
\simeq
\frac{f_\pi^*}{f_\pi}.
\label{brown-rho}
\end{equation}

Since he recognized that many-body effects are important, his point was that such medium modifications should be viewed as an effective starting point rather than as a direct modification of the fundamental parameters of QCD. Such many-body effects naturally induce width broadening in the medium, which is found to be particularly important for the $\rho$ meson at finite temperature.

\section{Refinements of Brown–Rho Scaling}

It is important to note that, despite the criticisms and limitations of its
original formulation, Brown--Rho scaling played an important role in advancing
our understanding of medium modifications of hadrons and stimulated extensive
subsequent studies, both theoretical and experimental. Let us now turn to what
we have learned since then and how Brown--Rho scaling can be naturally refined
and reinterpreted in light of these developments.

\subsection{Quark versus gluon condensates}

One important aspect that Mannque recognized and incorporated into his later
work was the need to distinguish the medium dependence of the quark and gluon
condensates.

The early lattice studies of finite-temperature QCD provided an important
starting point for thinking about the relation between chiral symmetry
restoration and deconfinement. In an early lattice calculation,
Kogut et al.~\cite{Kogut:1982rt} studied $SU(2)$ and $SU(3)$ gauge theories
using the quenched approximation and compared the deconfinement transition
with the restoration of chiral symmetry. For $SU(3)$ they found the two
transition temperatures to be remarkably close,
\begin{align}
    1 \leq \frac{T_\chi}{T_c} \leq 1.05,
\end{align}
with $T_c$ of order $200$ MeV. This result naturally suggested that the
loss of confinement and the restoration of chiral symmetry might be two
manifestations of the same change in the QCD vacuum. Such an observation may
have encouraged the identification of the quark and gluon condensates implied
by Eq.~\eqref{chiraltogluon}. It is important, however, to remember that this
early calculation neglected dynamical fermion loops, and that subsequent
calculations did not immediately establish a universal coincidence of the two
transitions. Thus, the early lattice result should be regarded as evidence for
a close correspondence rather than as a proof that the two phenomena are
governed by the same order parameter.

A particularly suggestive observation came from the behavior of Wilson
loops above the transition. While the space-time Wilson loop changes from an
area law to a perimeter law across the deconfinement transition, indicating
the loss of the usual temporal confinement of a static quark-antiquark pair,
the space-space Wilson loop continues to exhibit an area law above
$T_c$~\cite{Manousakis:1986jh}. 
This observation led to the question of whether the gluon condensate itself
should disappear at the deconfinement transition. In my work,
``Gluon Condensates above $T_c$''~\cite{Lee:1989qj}, I calculated the electric
and magnetic gluon condensates in $SU(2)$ lattice gauge theory above $T_c$.
The important result was that both condensates remain nonzero above the
transition. The condensates in this calculation were defined after removing
the perturbative contribution. Thus, the calculation provided an early
indication that the nonperturbative gluonic vacuum is not simply destroyed at
$T_c$, and that the gluon condensate should not be regarded as an order
parameter for chiral symmetry restoration.
Later, the electric and magnetic gluon condensates were extracted separately
from lattice thermodynamic quantities such as the energy density and
pressure~\cite{Lee:2008xp}. These analyses further emphasized that the
temperature dependence of the gluonic sector is not exhausted by the behavior
of the quark condensate.

There is also a useful way to see this distinction directly from the
spectrum of the Dirac operator. Because the static heavy quark probes the
gluonic gauge configurations, the gluon condensate can also be expressed in
terms of the eigenvalue density of the Dirac operator~\cite{Lee:2013es},
\begin{align}
    \left\langle
    \frac{\alpha_s}{\pi}G^2
    \right\rangle
    =
    12\left\langle
    \sum_{\lambda}\rho(\lambda)
    \right\rangle .
    \label{gluon_spectral}
\end{align}
Here the sum denotes the corresponding spectral integral. In contrast, the
Banks--Casher relation gives
\begin{align}
    \langle\bar q q\rangle=-\pi\rho(0).
    \label{bc_condensate}
\end{align}
Thus, while the chiral condensate is determined directly by the density of
Dirac eigenvalues at the origin, the gluon condensate receives contributions
from the full eigenvalue spectrum. In particular, the contribution from the
infinitesimal region around $\lambda=0$ is proportional to
$\rho(0)d\lambda$ and vanishes as $d\lambda\rightarrow0$. Consequently, the
physics associated with the chiral order parameter and that associated with
the gluon condensate need not have the same spectral origin
~\cite{Lee:2013es}. This provides a QCD-based reason why the identification
in Eq.~\eqref{chiraltogluon} cannot in general be regarded as a fundamental
relation.

The persistence of a substantial gluon condensate above $T_c$ therefore
suggests separating the gluonic contribution to the QCD trace anomaly into
components with different behavior under chiral restoration. This idea was
subsequently developed by Mannque and collaborators in terms of ``soft
glue'' and ``hard glue,'' or ``epoxy.'' The soft component is associated with
the part of the gluonic vacuum structure that changes with chiral restoration,
whereas the hard component remains substantial above $T_c$ and represents
the residual explicit breaking of scale invariance. In this picture one may
schematically write
\begin{align}
    \langle G^2\rangle
    =
    \langle G^2\rangle_{\rm soft}
    +
    \langle G^2\rangle_{\rm hard},
\end{align}
with
\begin{align}
    \langle G^2\rangle_{\rm soft}
    &\longrightarrow 0
    \qquad\text{near chiral restoration},
    \\
    \langle G^2\rangle_{\rm hard}
    &\neq 0
    \qquad\text{above }T_c .
\end{align}
Mannque later formulated this effect explicitly in effective theories with
separate soft and hard dilaton degrees of freedom~\cite{Lee:2009rz}.

This development represents an important refinement of the original
Brown--Rho picture. The medium dependence of hadronic parameters should not
be tied indiscriminately to the full gluon condensate, since only part of the
gluonic scale-breaking dynamics is associated with the spontaneous breaking
of chiral symmetry. The remaining hard component of the gluon condensate can
survive chiral restoration and need not participate in the same scaling of
hadronic parameters.

Thus, rather than identifying quark condensate to the dilaton field as in Eq.~\eqref{chiraltogluon}, one should distinguish the chiral order parameter from the gluonic
contribution to the trace anomaly. This distinction provides a more natural
interpretation of the medium dependence entering an effective hadronic
Lagrangian and forms an important part of the subsequent refinement of
Brown--Rho scaling.  Furthermore, the simple relation given in Eq.~\eqref{brown-rho} will not hold anymore.

\subsection{Vector mesons}

It is important to note that, despite the criticisms and limitations of its original formulation, Brown--Rho scaling played an important role in advancing our understanding of medium modifications of hadrons and stimulated extensive subsequent studies, both theoretical and experimental. Let us now turn to what we have learned since then and how Brown--Rho scaling can be naturally refined and reinterpreted in light of these developments.

One of the important questions concerns the identification of the in-medium hadron mass with the chiral order parameter. The original Brown--Rho argument associated the scaling of the vector-meson mass with that of the pion decay constant within the hidden-local-symmetry formulation. This identification requires additional assumptions about the medium dependence of the parameters of the effective theory and is therefore not, by itself, a consequence of chiral symmetry. In particular, there is no general symmetry argument requiring the vector-meson mass to vanish when the chiral order parameter vanishes.

A strict consequence of chiral symmetry restoration is the degeneracy of chiral partners. For the vector and axial-vector mesons, this requires
\begin{align}
m_\rho^* \longrightarrow m_{a_1}^*,
\end{align}
but does not require their common mass to approach zero. This distinction is important. Chiral symmetry constrains the difference between the vector and axial-vector channels, whereas the common mass of the two chiral partners can contain contributions that are invariant under chiral transformations.

This point can also be seen in the massive Yang--Mills formulation of the vector and axial-vector mesons. In this framework, the mass structure can be organized schematically such that the chiral-symmetry-breaking contribution enters the splitting between the vector and axial-vector states,
\begin{align}
m_{a_1}^2-m_\rho^2
\sim g^2 f_\pi^2,
\end{align}
up to convention-dependent factors. Thus, as the chiral order parameter decreases, the natural consequence is a reduction of the $\rho$--$a_1$ mass splitting. The common mass need not disappear with $f_\pi$. In this respect, the massive Yang--Mills formulation provides a useful perspective on the distinction between the chiral-symmetry-breaking and chirally symmetric contributions to hadron masses. In our earlier work on vector-meson properties in a low-temperature pion gas, we studied the hidden-gauge and massive Yang--Mills formulations on the same footing and showed that both formulations satisfy the low-temperature chiral theorems for the vector and axial-vector currents~\cite{Lee:1994wx}. The analysis also demonstrated that the medium modification of the $\rho$ mass is constrained by the chiral structure of the theory rather than being determined solely by a replacement $f_\pi\to f_\pi^*$.

This observation is supported by a QCD-based analysis of the origin of hadron masses. In the first QCD sum rule approach for vector mesond in medium\cite{Hatsuda:1991ez}, the four-quark condensate were not separated into chiral symmetry breaking and symmetric part. However in  Ref.~\cite{Kim:2020zae,Kim:2021xyp} we were able to  separate the contributions of operators associated with chiral-symmetry breaking from those that remain in the chirally symmetric limit. Applying the sum rules to both $\rho$ and $a_1$, we were able to determine the symmetric and breaking part separately. We then studied the hadron masses after removing the chiral-symmetry-breaking contributions while retaining the chirally symmetric contributions at their vacuum values. For the $\rho$ and $a_1$, the two states become degenerate in the chiral-symmetry-restored vacuum, with a common mass of approximately
\begin{align}
m_{\rho}^{\rm sym}\simeq m_{a_1}^{\rm sym}
\simeq 570~{\rm MeV}.
\end{align}
Thus, while the chiral-symmetry-breaking contribution is essential for generating the physical mass splitting between the $\rho$ and $a_1$, a substantial part of the vector-meson mass remains even after the chiral-symmetry-breaking contributions are removed. More generally, our analysis found that the chiral partners $(\rho,a_1)$ and $(K^*,K_1)$ become degenerate with masses in the range of approximately $500$--$600$ MeV in the chiral-symmetry-restored vacuum.

This result suggests a useful refinement of the original Brown--Rho interpretation. Rather than identifying the entire hadron mass with the chiral order parameter, one should separate schematically
\begin{align}
m_H^*
=
m_H^{\rm sym}
+
\Delta m_H^{\chi},
\end{align}
where $m_H^{\rm sym}$ denotes the chirally symmetric contribution and $\Delta m_H^{\chi}$ denotes the contribution associated with spontaneous chiral symmetry breaking. For a pair of chiral partners, the physically relevant quantity is therefore their mass splitting. One should therefore replace Eq.~\eqref{brown-rho} as 
\begin{align}
\frac{m_{a_1}^{*\,2}-m_\rho^{*\,2}}
     {m_{a_1}^{2}-m_\rho^{2}}
\sim
\frac{m_{K_1}^{*\,2}-m_{K^*}^{*\,2}}
     {m_{K_1}^{2}-m_{K^*}^{2}}
\sim
\left(\frac{f_\pi^*}{f_\pi}\right)^2.
\label{chiralpartnerscaling}
\end{align}
Its significance is that the chiral order parameter is  associated with the splitting between chiral partners than with their common mass.

In the next section, I will discuss why I did not include similar relation between parity partners for the baryon states.

\subsection{$U_A(1)$ effects}

The distinction between ordinary chiral symmetry breaking and the breaking of
$U_A(1)$ becomes particularly important when considering topologically
non-trivial gauge configurations\cite{Lee:1996zy}. The two effects need not scale in the same
way. The ordinary quark condensate is controlled, in the thermodynamic limit,
by the accumulation of near-zero eigenmodes of the Dirac operator, whereas the effects of $U_A(1)$-breaking effects are due to exact zero modes.  

The near zero modes are related to the chiral order parameter through the
Banks--Casher relation,
\begin{align}
    \langle\bar q q\rangle=-\pi\rho(0),
\end{align}
where $\rho(0)$ denotes the density of near-zero Dirac. This is the order parameter associated with the spontaneous breaking
of the non-singlet chiral symmetry,
\begin{align}
    SU(N_f)_L\times SU(N_f)_R
    \longrightarrow SU(N_f)_V.
\end{align}
It should not, however, be identified with the contribution from exact
topological zero modes.

The distinction can be seen directly in the functional integral. After
integrating out the quarks, the QCD partition function can be written as
\begin{align}
    Z[m]
    =
    \int {\cal D}A\,
    e^{-S_{\rm YM}}
    \det(\slashed D+m)^{N_f}.
\end{align}
In a sector with nonzero topological charge, the exact zero modes can be
separated from the nonzero-mode determinant. For example, in the
$\nu=1$ sector one may write schematically
\begin{align}
    Z[m]_{\nu=1}
    =
    \int {\cal D}A_{\nu=1}\,
    e^{-S_{\rm YM}}\,
    {\rm Det}'(\slashed D+m)\,
    \det_{sf}
    \left(
      \int d^4x\,
      \bar\psi_0(x) m \psi_0(x)
    \right),
\end{align}
where ${\rm Det}'$ denotes the determinant with the exact zero modes
removed. The index theorem fixes the difference between the numbers of
right- and left-handed zero modes,
\begin{align}
    \nu=n_+-n_-.
\end{align}
The important point is that
the exact zero-mode contribution carries explicit powers of the quark mass,
whose power depends on the number of zero modes and flavors.

For the ordinary quark condensate, the contribution from a non-trivial
topological sector is therefore suppressed in the chiral limit. For
$N_f=2$, for example, the $\nu=\pm1$ contribution to $\langle\bar q q\rangle$
is of order $m$, while for general $N_f$ it starts at
$O(m^{N_f-1})$~\cite{Lee:1996zy}. Consequently, in the chiral limit the
Banks--Casher relation is saturated by the near-zero spectrum in the
topologically trivial sector. This does not mean that the topologically
non-trivial sectors are irrelevant. As emphasized in Ref.~\cite{Lee:1996zy},
the exact zero modes in $\nu\neq0$ sectors can contribute to higher correlation functions that survive the chiral limit.

This distinction becomes especially transparent in correlation functions
that are related by both a non-singlet chiral rotation and a $U_A(1)$
rotation. For $N_f=2$, consider, for example, the difference between the
scalar-isoscalar and scalar-isovector correlators,
\begin{align}
    \Delta_{\sigma\delta}(x)
    &=
    \left\langle
      \bar q(x)q(x)\,
      \bar q(0)q(0)
    \right\rangle
    -
    \left\langle
      \bar q(x)\tau^3q(x)\,
      \bar q(0)\tau^3q(0)
    \right\rangle .
\end{align}
In the chirally restored phase, the contribution associated with the
near-zero spectrum vanishes with $\rho(0)$. Nevertheless, the topologically
non-trivial sector contains exact zero modes which can saturate the external
fermion fields. The resulting contribution to the difference survives in the
chiral limit. In the two-flavor theory this is the familiar $U_A(1)$-violating
effect associated with the 't Hooft interaction,
\begin{align}
    {\cal L}_{\rm tHooft}
    \sim
    \det_{f,f'}
    \left(\bar q_{Rf}q_{Lf'}\right)
    +{\rm h.c.},
\end{align}
or explicitly,
\begin{align}
    {\cal L}_{\rm tHooft}
    \sim&
    \left(\bar u_Ru_L\right)
    \left(\bar d_Rd_L\right)
    -
    \left(\bar u_Rd_L\right)
    \left(\bar d_Ru_L\right)
    +{\rm h.c.}
\end{align}
This operator preserves the non-singlet chiral symmetry but violates
$U_A(1)$, and therefore it can distinguish correlation functions that become
identical when only the non-singlet chiral symmetry is restored. In terms of meson fields, if the strength of the 't Hooft interaction is $K$, its contribution to the meson masses are as follows \cite{tHooft:1986ooh}
\begin{align}
    V_{U_A(1)}=-K \bigg(\sigma^2-\eta^2-a^2+\pi^2 \bigg).
\end{align}
Hence, while $K$ cancels out in the $\pi $-$\sigma$ mass difference, it contributes to the mass difference between the $\sigma$ and $a$ mesons.

This observation was the main point of our analysis with Hatsuda in
Ref.~\cite{Lee:1996zy}. We showed that the restoration of
$SU(N_f)_L\times SU(N_f)_R$ does not by itself guarantee the complete
restoration of $U_A(1)$ in the hadron spectrum. For $N_f=2$, topologically
non-trivial configurations can give a finite contribution to appropriate
two-point functions even in the chirally restored phase. For $N_f\geq3$,
the corresponding two-point contribution is suppressed by additional powers
of the quark masses, although in the real world the finite strange-quark
mass can make such effects non-negligible. More generally, the topological
contribution to higher-point correlation functions can survive whenever the
external fermion fields are sufficient to saturate the zero modes
~\cite{Lee:1996zy}.

This provides a useful perspective for hadron spectroscopy. The contribution
of the near-zero spectrum and the contribution associated with exact
topological zero modes need not have the same dependence on the quark mass,
volume, or temperature. Schematically, one may therefore distinguish
\begin{align}
    \Delta {\cal C}
    =
    \Delta {\cal C}_{\rm near\mbox{-}zero}
    +
    \Delta {\cal C}_{\rm top},
\end{align}
where the first term is associated with the accumulation of near-zero
nonzero modes and the second with the topologically non-trivial sector.
The former is directly connected to the conventional chiral order parameter,
whereas the latter can remain finite in $U_A(1)$-sensitive correlation
functions even after the ordinary chiral condensate has vanished.

This distinction may be particularly relevant for parity-partner states.
For a baryon correlator, which contains more fermion fields than a mesonic
two-point function, the exact zero modes can in principle be saturated in
ways that are different for different interpolating currents. Consequently,
even after the near-zero-mode contribution associated with
$\langle\bar q q\rangle$ is suppressed by chiral restoration, a residual
topological contribution could distinguish correlation functions associated
with opposite-parity baryons if the corresponding currents are sensitive to
the $U_A(1)$-violating interaction. In such a situation one could schematically
write
\begin{align}
    M_{B_+}-M_{B_-}
    =
    \Delta M_B^{\rm chiral}
    +
    \Delta M_B^{U_A(1)}
    +\cdots ,
\end{align}
with
\begin{align}
    \Delta M_B^{\rm chiral}\longrightarrow0
\end{align}
when the non-singlet chiral symmetry is restored, while
$\Delta M_B^{U_A(1)}$ need not vanish if topological zero-mode effects
remain. This should be regarded as a possible mechanism rather than a
model-independent prediction: whether the topological contribution actually
survives in a given baryon channel depends on the quantum numbers and the
zero-mode saturation properties of the interpolating current.

A similar consideration applies to mesons. If two mesonic channels are
related by a $U_A(1)$ transformation in addition to the non-singlet chiral
symmetry, their correlators can remain different after the ordinary chiral
order parameter has vanished. The scalar--isovector and
pseudoscalar--isosinglet channels provide the classic example. The same
logic applies more generally to channels whose difference is sensitive to
the $U_A(1)$ anomaly. By contrast, for channels related only by
$SU(N_f)_L\times SU(N_f)_R$, restoration of the non-singlet chiral symmetry
is sufficient to enforce the corresponding degeneracy, provided other
explicit symmetry-breaking effects are absent.

The important lesson is therefore that the disappearance of
$\langle\bar q q\rangle$ should not be taken as synonymous with the
disappearance of all anomalous effects. The quark condensate measures the
extensive accumulation of near-zero Dirac modes, whereas exact topological
zero modes can leave their characteristic imprint on higher correlation
functions through the $U_A(1)$-violating interaction. In this sense, the
temperature or density dependence of $U_A(1)$-sensitive observables can have
a qualitatively different scaling from that of the ordinary chiral order
parameter. This provides a natural framework in which parity partners, or
mesonic channels distinguished specifically by $U_A(1)$, can retain a
splitting even when the conventional chiral symmetry has been restored.
A quantitative realization of this idea requires the topological
contribution to the relevant hadronic correlator to be calculated explicitly.

The difference between exact zero modes and near-zero modes should therefore be carefully treated in a realistic lattice calculation. The exact zero modes associated with topologically non-trivial configurations are not necessarily easy to sample and identify accurately in a lattice calculation. Therefore, extracting the $U_A(1)$ breaking effect from generated gauge-field configurations and the corresponding quark propagator should be done with extreme care. Otherwise, an apparent restoration of $U_A(1)$ symmetry when chiral symmetry is restored in the lattice calculation could potentially be an artifact of an inadequate treatment of the exact zero modes. Thus, a lattice analysis of $U_A(1)$ restoration needs to distinguish carefully between exact zero modes, topological near-zero modes, and the remaining near-zero spectrum.  For these reasons,  I will not include the baryons or their mass difference in the refined Brown-Rho scaling.

\section{Summary}

Mannque spent three months at Yonsei University as a visiting professor, where he gave a special lecture series on renormalization-group methods, Fermi-liquid theory, and hidden local symmetry.
Although we never coauthored a paper, we remained in continuous communication from 1984 until his final days. My career would have been much less interesting without his constant challenges, provocative questions, and brilliant ideas. I dedicate this paper to his memory, with gratitude for a friendship and intellectual exchange that lasted more than four decades.

\section{Acknowledgment}




\end{document}